\documentstyle[twocolumn,prl,aps]{revtex}
\def\lsim{\mathrel{\rlap{
\lower4pt\hbox{\hskip-3pt$\sim$}}
    \raise1pt\hbox{$<$}}}     %less than or approx. symbol
\def\gsim{\mathrel{\rlap{
\lower4pt\hbox{\hskip1pt$\sim$}}
    \raise1pt\hbox{$>$}}}     %greater than or approx. symbol

\def\beq{\begin{eqnarray}}
\def\eeq{\end{eqnarray}}
\newcommand{\be}{\begin{eqnarray}}
\newcommand{\ee}{\end{eqnarray}}

\def\np{Nucl. Phys.}
\def\pr{Phys. Rev.}
\def\prl{Phys. Rev. Lett.}
\def\pl{Phys. Lett.}
\def\del{\partial}
\def\Tr{{\rm Tr}}
\def\L{{\cal L}}
\def\Linv{{\cal L}_{inv}}
\def\Lsb{{\cal L}_{sb}}
\def\la{\langle}
\def\ra{\rangle}
\def\der{\mbox{d}}

\begin{document}                                                              
\draft
\title{Connection between Effective Chiral Lagrangians and Landau-Migdal
Fermi Liquid Theory of Nuclear Matter}
\author{Mannque Rho}
\address{
C.E.A.-- Saclay, Service de Physique Th\'eorique,  F-91191 Gif-sur-Yvette Cedex,
France}
\date{\today}
\maketitle
\begin{abstract}

This lecture is based in part on work done in collaboration with G.E. Brown 
and on a recent paper co-authored with Bengt Friman. It deals 
with making a connection between effective chiral Lagrangians -- low-energy
effective theory of QCD -- and Landau Fermi liquid theory extended by
Migdal to nuclear matter. I discuss how to obtain
a link between observables in relativistic heavy-ion processes
and low-energy spectroscopic data, giving a new insight into how
chiral dynamics manifests itself in nuclear systems.
\end{abstract}
\begin{narrowtext}

\section{Introduction}
\indent
It is generally accepted that the physics of low-energy strong interactions
relevant for nuclear physics is governed by effective theories based on chiral
Lagrangians. At long-wavelength limit, the strategy of implementing chiral
Lagrangian field theory is in principle known and has been successfully
applied to low-energy interactions in the pseudo-Goldstone (i.e., pion)
sector. Now how to implement such a strategy in nuclear dynamics that
involves many-body interactions is an entirely different matter and 
has met with little success. This is an important issue however for 
ultimately understanding what happens in relativistic heavy-ion physics
looking for extreme states of hadronic matter since the experiments
looking for such states will sample all ranges of strong interactions,
from low to high energy and hence would require highly nonperturbative
to perturbative regimes of QCD. This means that the theory will have to 
explain correctly low-energy nuclear processes in order to make sense
in the extreme conditions probed in relativistic heavy-ion collisions.

In this lecture I would like to describe some of the recent developments
along this direction \cite{br91,br95,brPR95,frimanrho}.

Let me start with what I consider to be an exciting new development in
nuclear physics. It was shown in
recent publications by Li, Ko and Brown \cite{lkb95} that the dilepton
production data of CERES \cite{CERES} and HELIOS-3 \cite{HELIO}
can be simply and quantitatively understood if the mass
of the vector mesons $\rho$ and $\omega$ scales in dense and/or hot medium
according to the scaling (BR scaling)
proposed by Brown and Rho \cite{br91}. That
the vector mesons ``shed" their masses as the density (or temperature)
of the matter increases is expected in an intuitive interpretation of
the interplay of the condensation of quark-antiquark pairs
and the dynamical generation of light-quark hadron masses
and is in fact corroborated by QCD sum rules \cite{sumrules,Jin1}
and model calculations \cite{model}. Thus, the dilepton data are consistent
with the most conspicuous prediction of BR scaling
(see \cite{wambach} for other mechanisms). 
The proposal of \cite{br91}, however, goes further than this and
makes a statement on the relation between the scaling of
meson masses and that of baryon masses:
\be
\frac{m_M^\star}{m_M}\approx \sqrt{\frac{g_A}{g_A^\star}}
\frac{m_B^\star}{m_B}\approx \frac{f_\pi^\star}{f_\pi}\equiv
\Phi (\rho)\label{br}
\ee
where the subscript $M$ stands for light-quark non-Goldstone mesons,
$B$ for light-quark baryons, $g_A$ the axial-current coupling
constant and $f_\pi$ the pion decay constant.
The star denotes an in-medium quantity. (Although temperature
effects can also be discussed in a similar way, we will
be primarily interested in density effects in this paper.)

Two immediate questions are raised in these developments:
Firstly, is there evidence that the baryon mass scaling and the
meson mass scaling are related as implied by the chiral Lagrangian?
Secondly, we know from the Walecka model of nuclear matter \cite{waleckamodel}
that the ``scalar mass" of the nucleon drops as a function of density
and that this reduction of the nucleon mass has significant consequences on
nuclear spectroscopy and the static properties of nuclei. The question is:
Is BR scaling related to the ``conventional" mechanism for the reduction of
the nucleon mass in nuclear matter and if so, how does it manifest itself in
low-energy nuclear properties? Put differently, can a single chiral Lagrangian
explain at the same time low-energy nuclear processes studied in a conventional
way since a long time and the high-energy processes to be probed in heavy-ion
collisions?

The aim of this lecture is to show, based on
recent work \cite{frimanrho,FR,br95}, that the connection between the
meson and baryon scalings can be
made using the Landau-Migdal theory of nuclei and nuclear matter.
See \cite{GEB95} for a related discussion. Our starting
point is the effective chiral Lagrangian used in \cite{br91} where
the scale anomaly of QCD is incorporated
and baryons arise as skyrmions. This theory is mapped onto an
effective meson-baryon chiral Lagrangian. We establish the relation between
chiral and Walecka mean fields in medium as suggested in \cite{br95} and then
invoke the Galilei invariance argument of Landau, which relates the nucleon
effective mass to the Landau Fermi liquid parameters. Thus, we establish a
relation between the parameters in eq.~(\ref{br}) and the Landau parameters. We
discuss how this relation can be tested with the effective $g_A^\star$ and
the gyromagnetic ratios $\delta g_l$ in nuclear matter. This then supplies
a novel relation between the scaled
masses, which may be reflected in the spectrum of dileptons produced in
relativistic heavy-ion collisions, and low-energy spectroscopic information,
$g_A^\star$ and $\delta g_l$. It also supplies an indirect and nontrivial
connection between quantities figuring in chiral Lagrangians of
QCD and those appearing in familiar many-body theory. 

In order to avoid unnecessary complications we shall use the
nonrelativistic approach to Landau Fermi liquid theory, referring to
results obtained in the relativistic formulation \cite{baymchin,BWBS}
where appropriate.
\section{Renormalization Group Fixed Points and Chiral Lagrangians}
\indent
Before I enter into the main topic of this lecture, let me 
mention one important issue here.
Given an effective Lagrangian that describes QCD in the low energy 
nonperturbative regime, how does one use it to describe nuclear
many-body processes? The answer to this question is not known
at the moment. In fact it is not even clear whether it is meaningful
or necessary to ask such a question.  Nonetheless some of us have been
asking this question since some time.

Let me describe briefly how I understand this problem.

Let me start by assuming that Walecka mean field theory of nuclear
matter \cite{waleckamodel} is correct. Now to go from chiral Lagrangians
to Walecka theory, thus far, two approaches have been proposed.
One approach is to look at nuclear matter as a ``chiral liquid"
resulting as a solitonic matter from an effective chiral action \cite{lynn}.
I do not have much to add to this approach; as it stands, no quantitative
insight can be gained from it. Instead I will focus on the second approach
which is to use a chiral Lagrangian endowed with four-Fermi interactions
having the quantum numbers of an isoscalar vector (say, $\omega$) and 
isoscalar scalar (say, $\sigma$) exchanges treated {\it in mean field 
with BR scaling} \cite{br95,gelmini,pmr95}. For $N=Z$ nuclear matter,
pions average out (see later however for their important role).
I will take this as an (heuristic) evidence that effective chiral 
Lagrangians can be mapped to Walecka theory. The mean fields provide
the Fermi surface in the usual way. As was shown in \cite{frimanrho} and 
will be discussed shortly, this theory can then be connected to
Landau Fermi liquid theory of nuclear matter \cite{landau,migdal}.

Now given the effective Lagrangian with the Fermi surface, one can
then apply the renormalization group arguments developed in
condensed matter physics \cite{shankarpol} in the following way.
In the presence of a Fermi surface, the four-Fermi interactions 
yield the Landau parameters of quasiparticle interactions as
renormalization group fixed points in the sense that the 
four-Fermi interactions are marginal, i.e.,  with 
vanishing $\beta$ functions. As in condensed matter systems,
one might have ``BCS"-type four-Fermi interactions that can become
marginally relevant (e.g., kaon condensation) but here I will
not deal with this possibility.

There is one basic difference in nuclear matter
from condensed matter systems and that
is that in QCD at quantum level, there is a scale anomaly. This
scale anomaly associated with the fact that the energy-momentum
tensor has a nonvanishing trace in QCD at quantum level will give
rise to an anomalous dimension to the scalar field associated with
the trace anomaly. This suggests that an effective chiral Lagrangian
with the scale anomaly incorporated (as used in BR scaling) can be
mapped to Walecka-type theory with the scalar field $\sigma$
having an anomalous dimension. I believe that this is the
mechanism behind the successful mean-field chiral Lagrangian
theory of nuclei and nuclear matter by Furnstahl et al \cite{tang} who find
phenomenologically a large anomalous dimension $d_a\approx 1.7$
for the $\sigma$ field for which the incompressibility modulus
comes out to be $K\sim 200$ MeV and many-body forces are strongly suppressed
at the saturation point \cite{song}. This turns out to be what is needed
for a mean field theory of the chiral phase transition proposed in
\cite{bbr96}.

In the rest of the lecture, I discuss how Landau theory can be 
incorporated into this chiral Lagrangian scheme.

\section{BR Scaling}
\indent
The BR scaling relation (\ref{br}) that relates the dropping of light-quark
non-Goldstone-boson masses to that of the nucleon mass which in
turn is related to that of the pion decay constant was first derived
by incorporating the trace anomaly of QCD
into an effective chiral Lagrangian. The basic idea can be summarized as
follows.  We wish to write an effective chiral Lagrangian which at
mean-field level reproduces the quantum trace anomaly while including
higher chiral order effects relevant for nuclear dynamics. To do this,
we write the effective Lagrangian in two parts
\be
\L=\Linv +\Lsb \label{twopieces}
\ee
where $\Linv$ is the scale-invariant part and $\Lsb$ the
scale-breaking part of the effective Lagrangian. We introduce the {\it
chiral-singlet} scalar field $\chi$, as an interpolating field for
$\Tr\ G^2$, \be \theta_\mu^\mu=\frac{\beta (g)}{2g} \Tr\,
G_{\mu\nu}G^{\mu\nu}\equiv \chi^4, \ee where we have dropped the quark
mass term (here we consider the chiral limit). The simplest possible 
invariant piece of the Lagrangian then takes the form
\be
\Linv&=&\frac{f_\pi}{4} \left(\frac{\chi^2}{\chi_0^2}\right)\Tr\, (\del_\mu U
\del^\mu U^\dagger)\nonumber\\
&& + \frac{1}{32  g^2}\Tr\, [U^\dagger\del_\mu U,
U^\dagger\del_\nu U]^2 +\cdots\label{effectiveL}
\ee
where $\chi_0$ is a number which we define to be the expectation value
of $\chi$ in matter-free vacuum and
the ellipsis stands for other-scale invariant terms including
the kinetic energy term for the $\chi$ field. Note that this is the simplest
possible form based on the most economical assumption.  One could perhaps write
much more complicated and yet scale-invariant forms using the same
set of fields but invoking different assumptions, and thus obtain a 
different type of scaling. Experiments will tell us which one is the right
form.

As for the scale-breaking term $\Lsb$, we assume that it
contains just the terms needed to reproduce the full trace anomaly. We 
add other scale-invariant terms representing higher chiral order terms to
assure the correct vacuum potential which we shall call $V(\chi,U)$.
Fortunately all we need to know about the potential $V$ is that it
contains a source for the $\chi$ mass term and that, for a given density, it
attains its minimum at $\chi^\star=\la \chi \ra^\star$ in
the sense of the Coleman-Weinberg mechanism \cite{coleman}. (We will return
later to what this quantity $\chi^\star$ represents physically.)

The fact that the vacuum expectation value is obtained by minimizing the
potential, which contains a scale-breaking term, implies that we are
treating the breaking of the scale invariance as a spontaneous symmetry
breaking. It is well-known that the spontaneous breaking of the scale
symmetry occurs only if it is explicitly broken, since otherwise the
potential would be flat \cite{zumino}. Given the ground state
characterized by $\chi^\star$ which is fixed by the anomaly, we then 
shift the field in (\ref{twopieces}) 
\be
\chi (x)=\chi^\prime (x) +\chi^\star.
\ee
After shifting, we still have the scale-invariant and scale-breaking
pieces although the manifest invariance is lost as is the case with {\it all}
spontaneously broken symmetries. The low-energy physics for the scaling
we are interested in is lodged in the former. Since the theory contains two
parameters, $f_\pi$ and $g$, we define
\be
f_\pi^\star&=&f_\pi\frac{\chi^\star}{\chi_0},\nonumber\\
g^\star &=& g.
\ee
The second relation follows since the Skyrme quartic term in (\ref{effectiveL})
is scale-invariant by itself. I will argue later that
in the baryon sector
there is an important radiative correction -- absent in the meson sector --
which modifies this scaling behavior.
This allows us to redefine the parameters that appear in the chiral
Lagrangian in terms of the ``starred" parameters $f_\pi^\star$ and
$g^\star$. Since the KSRF relation \cite{KSRF} is an exact low-energy theorem
as shown by Harada, Kugo and Yamawaki \cite{harada}, it is reasonable to
assume that it holds also in medium. This
leads to
\be
m_V^\star/m_V\approx \frac{f_\pi^\star g^\star}{f_\pi g}\approx
\frac{f_\pi^\star}{f_\pi}
\equiv\Phi (\rho) <1\ \ \ {\rm for}\ \ \rho\neq 0
\ee
where the subscript V stands for $\rho$ or $\omega$ meson.
Similarly the mass of the scalar field is reduced
\be
m_\sigma^\star/m_\sigma\approx \Phi (\rho).
\ee
Here we denote the relevant scalar field by the usual notation
$\sigma$ for reasons given below.

Now in order to find the scaling behavior of the nucleon mass, we use the
fact that the nucleon arises as a soliton (skyrmion) from the effective
chiral Lagrangian as in the free-space. The soliton mass goes like
\be\label{skyrmion}
m_S\sim f_\pi/g.
\ee
If one assumes that by the same token the coupling constant $g$ in the
soliton sector is not modified in the medium, eq.~(\ref{skyrmion}) implies
that the nucleon mass is also proportional to $f_\pi^\star$,
\be
m_N^\star/m_N\sim  \Phi (\rho).
\ee
However there is a caveat to this. When it comes to the
nucleon effective mass, there is one important non-mean-field effect
of short range that is known to be important. This is an
intrinsically quantum effect that cannot be accounted for in low orders of
the chiral expansion, namely the mechanism that quenches the axial-current
coupling constant $g_A$ in nuclear matter.
This effect is closely related to the Landau-Migdal interaction in the
spin-isospin channel $g_0^\prime$ (involving $\Delta$-hole excitations)
as discussed in \cite{brPR95,delta}. The
axial-vector coupling constant of the skyrmion is governed by
coefficient $g$ of the Skyrme quartic term. This implies that
in the baryon sector, the mean-field argument, which is valid in the mesonic
sector, needs to be modified. This is reminiscent of the deviation in the
nucleon electromagnetic form factor from the vector dominance model which
works very well for non-anomalous processes involving mesons.
These two phenomena may be related.

As shown in \cite{br91,mr88}, a more accurate expression, at least
for densities up to $\rho\sim \rho_0$, is\footnote{This was
derived using the scaling behavior of the Skyrme quartic
Lagrangian and the relation between $g_A$ and the coefficient $g$.
Although this relation is justified strictly at the large $N_c$ limit
(where $N_c$ is the number of colors), we think
that it is generic and will emerge in
any chiral model that has the correct symmetries.}
\be
m_N^\star/m_N\approx \Phi (\rho)\sqrt{\frac{g_A^\star}{g_A}} .\label{Nmass}
\ee
This relation will be used later to deduce a formula for $g_A^\star$
in nuclear matter.
Beyond $\rho=\rho_0$, we expect that $g_A^\star$ remains constant
($g_A^\star = 1$) and that $\Phi$ scaling takes over
except near the chiral phase transition at which the coupling constant
$g$ will fall according to the ``vector limit." \cite{brPR95}

\subsection{What is $\chi^\star$ ?}
\indent
The $\chi$ field interpolating as $\chi^4$ for the dimension-4 field
$\Tr G_{\mu\nu}^2$ may be dominated by a scalar glueball field, which
perhaps could be identified with the $f_J (1710)$ seen in lattice
calculations \cite{lattice}. However, for the scaling we are
discussing which is an intrinsically low-energy property, this is too
high in energy scale. In the  
effective Lagrangian (\ref{effectiveL}), such a heavy degree of
freedom should not appear explicitly.  The only reasonable interpretation
is that the $\chi$ field has two components,
\be
\chi=\chi_h + \chi_l\label{sep}
\ee
corresponding to high (h) and low (l) mass excitations, and 
that the high mass (glueball) component $\chi_h$ is
integrated out.  The ``vacuum" expectation value we are interested in
is therefore $\la \chi_l \ra^\star$. The corresponding fluctuation
must interpolate $2\pi$, $4\pi$ etc. excitations as discussed in
\cite{brPR95} and it is this field denoted by $\sigma$ that becomes
the dilaton degenerate with the pion at the chiral phase transition as
suggested by Weinberg's mended symmetry \cite{mended}.  It is also
this component which plays an essential role in the relation between
chiral Lagrangians and the Walecka model \cite{gelmini,pmr95,br95}.
This procedure may also be justified by a phenomenological instanton
model anchored in QCD \cite{bbr96}.

For a more physical interpretation and a detailed discussion on the 
separation (\ref{sep}), see Adami and
Brown \cite{adamibrown}. A somewhat different separation is advocated
by Furnstahl et al. in \cite{tang}.

\subsection{Four-Fermi interactions}
\indent
In order to make contact with many-body theory of nuclear matter, we
reinterpret the BR scaling in terms of a
baryon chiral Lagrangian in the relativistic baryon
formalism. There is a problem with chiral counting in this formalism\footnote{
As we know from the work of Gasser, Sainio and Svarc \cite{gasser},
the relativistic
formulation of baryon chiral perturbation theory requires a special care
in assuring a correct chiral counting. What we will find below is that in
order to get to the correct formulation from the point of view of Landau
Fermi liquid theory of normal nuclear matter
and making contact with Walecka theory at mean-field
order, it is essential to keep relativistic corrections
from the start. This probably
has to do with the presence of the Fermi sea in the effective chiral
Lagrangian approach. This seems to suggest that the usual chiral counting
valid in free space needs to be modified in medium.} but
our argument will be made at mean-field order as in \cite{br95}.

The Lagrangian contains the usual pionic piece $\L_\pi$, the pion-baryon
interaction $\L_{N\pi}$ and the four-Fermi contact interactions
\be
\L_{4}=\sum_\alpha \frac{C_\alpha^2}{2} (\bar{N}\Gamma_\alpha N)
(\bar{N}\Gamma^\alpha N)
\ee
where the
$\Gamma^\alpha$'s are Lorentz covariant quantities -- including derivatives --
that have the
correct chiral properties. The leading chiral order  four-Fermi contact
interactions relevant for the scaling masses are of the form
\be
\L_{4}^{(\delta)}
=\frac{C_\sigma^2}{2} (\bar{N}N\bar{N}N) -\frac{C_\omega^2}{2}
(\bar{N}\gamma_\mu N\bar{N} \gamma^\mu N).\label{wform}
\ee
As indicated by our choice of notation, the first term can be thought of as
arising when a massive isoscalar scalar meson (say, $\sigma$) is integrated
out and similarly for the second term involving a massive isoscalar vector
meson (say, $\omega$). Consequently, we can make the identification
\be
C_\sigma^2=\frac{g_\sigma^2}{m_\sigma^2},\ \
C_\omega^2=\frac{g_\omega^2}{m_\omega^2}.\label{constant}
\ee
The four-Fermi interaction involving the $\rho$ meson quantum number will be
introduced below, when we consider the  electromagnetic currents.
As is well known \cite{gelmini,br95}, the first four-Fermi  interaction in
(\ref{wform}) shifts the nucleon mass in matter,
\be
m_N^\sigma=m_N-C_\sigma^2 \la \bar{N}N\ra. \label{sigmashift}
\ee
In \cite{br95} it was shown that this shifted nucleon mass scales the same
way as the vector and scalar mesons
\be \frac{m_V^\star}{m_V}\approx
\frac{m_\sigma^\star}{m_\sigma}\approx \frac{m_N^\sigma} {m_N}\approx  \Phi
(\rho).\label{universal}
\ee
This relation was referred to in \cite{br91} as ``universal scaling." There
are two points to note here: First as argued in \cite{br95}, the
vector-meson mass scaling applies also to the masses in (\ref{constant}).
Thus, in medium the meson mass should be replaced by $m_{\sigma,
\omega}^\star$. Consequently, the coupling strengths $C_\sigma$ and
$C_\omega$ are density-dependent.\footnote{I should point out that
for the purpose of the ensuing discussion, neither the detailed knowledge
of the ``heavy" degrees of freedom that give rise to the four-Fermi interactions
nor the specific form of the density dependence will be needed. What really
matters are the quantum numbers involved. The latter is invoked in reducing
various density-dependent parameters to the universal one, $\Phi (\rho)$.}
Second, the scaling can be understood in
terms of effects due to the four-Fermi interactions, which for nucleons on
the Fermi surface correspond to the
fixed-point interactions of Landau Fermi liquid theory according to Shankar
and Polchinski \cite{shankarpol}. We shall establish a direct connection to
the Landau parameters of the quasiparticle-interaction.

\section{Landau's Effective Mass of the Nucleon}
\indent
In the Landau-Migdal Fermi liquid theory of nuclear matter
\cite{landau,migdal}, the interaction between two
quasiparticles on the Fermi surface is of the form (neglecting tensor
interactions)
\be
{\cal F}(\vec{p},\vec{p}^\prime) &=& F(\cos \theta) + F^\prime (\cos
\theta)(\vec{\tau}\cdot\vec{\tau}^\prime)\nonumber\\
& +& G(\cos
\theta)(\vec{\sigma}\cdot\vec{\sigma}^\prime)
+ G^\prime (\cos \theta)
(\vec{\tau}\cdot\vec{\tau}^\prime)
(\vec{\sigma}\cdot\vec{\sigma}^\prime),
\eeq
where $\theta$ is the angle between $\vec{p}$ and
$\vec{p}^\prime$. The function $F(\cos \theta)$ can be expanded in
Legendre polynomials,
\beq
F(\cos \theta) = \sum_l F_l P_l(\cos \theta),
\eeq
with analogous expansions for the spin- and isospin-dependent
interactions. The coefficients $F_l$ etc. are the Landau Fermi liquid
parameters. Some of the parameters can be related to physical
properties of the system. The relation between the effective mass and
the Landau parameter $F_1$ (eq.~(\ref{mstar})) is crucial for our
discussion.\footnote{In this lecture, we make no use of the Landau
parameters $G$ and $G^\prime$. The  $G^\prime$ in particular is believed to
figure in the response to the weak axial current by nuclei. We shall see
that one can actually obtain a similar result using only the $F$ and
$F^\prime$ parameters in combination with the pionic interaction.}

%\begin{figure}[hbt]
%\setlength{\unitlength}{1mm}
%\begin{picture}(150,45)
%\put(60,8){\epsfig{file=opep.eps,
%height=35mm,angle=-90}}
%\end{picture}
%\hfill
%\caption{\label{opepfig} The
%one-pion-exchange interaction corresponding to the non-local
%four-Fermi term in the Lagrangian (\protect{\ref{4fermii}}).}\hfill
%\end{figure}

An important point of this paper is that one must distinguish between
the effective mass $m_N^\sigma$, which is of the same form as
Walecka's effective mass, and the Landau effective mass, which
is more directly related to nuclear observables.
To see what the precise relation is, we include the non-local
four-Fermi interaction due to the one-pion exchange term, $L^{(\pi)}_4$.

The total four-Fermi interaction that enters in the renormalization-group
flow consideration \`a la Shankar-Polchinski is then the sum
\be
\label{4fermii}
\L_4=\L_4^{(\pi)} +\L_4^{(\delta)}.
\ee
The point here is that the non-local one-pion-exchange term brings
additional contributions to the effective nucleon mass on top of the
universal scaling mass discussed above.
%\begin{figure}[t]
%\setlength{\unitlength}{1mm}
%\begin{picture}(150,45)
%\put(60,8){\epsfig{file=pifock.eps,
%height=40mm,angle=-90}}
%\end{picture}
%\hfill\caption{\label{fockfig} The one-pion exchange (Fock term)
%contribution to the nucleon self energy.}\hfill
%\end{figure}
\noindent
We now compute the nucleon effective mass with the chiral Lagrangian and
make contact with the results of Fermi liquid theory \cite{FR}.
We start with the single-nucleon energy in the non-relativistic
approximation\footnote{We treat the scalar and vector fields
self-consistently and the self-energy from
the pion exchange graph as a perturbation.}
\be
\epsilon (p) =\frac{p^2}{2 m_N^\sigma} +
C_\omega^2\la N^\dagger N\ra +\Sigma_\pi (p)
\label{energy}
\ee
where $\Sigma_\pi (p)$ is the self-energy from the pion-exchange Fock
term.
The self-energy contribution from the vector meson (second term on the
right hand side of (\ref{energy})) comes from an $\omega$ tadpole 
(or Hartree) graph.
The Landau effective mass $m_L^\star$ is related to the quasiparticle
velocity at the Fermi surface
\be
\label{velo}
\frac{\der}{\der p} \epsilon (p)|_{p=p_F}=\frac{p_F}{m_L^\star}
= \frac{p_F}{m_N^\sigma} +\frac{\der}{\der p}\Sigma_\pi (p)|_{p=p_F}.
\ee
Using Galilean invariance, Landau \cite{landau} derived a relation between the
effective mass of the quasi-particles and the velocity dependence of the
effective interaction described by the Fermi-liquid parameter $F_1$:
\be
\label{mstar}
\frac{m^\star_L}{m_N} = 1 + \frac{F_1}{3} = (1-\frac{\tilde{F_1}}{3})^{-1},
\ee
where $ \tilde{F_1} = (m_N/m^\star_L) F_1$.
The corresponding relation
for relativistic systems follows from Lorentz invariance and has been
derived by Baym and Chin \cite{baymchin}.

%\begin{figure}[t]
%\setlength{\unitlength}{1mm}
%\begin{picture}(150,45)
%\put(60,5){\epsfig{file=omega_hartree.eps,
%height=35mm,angle=-90}}
%\end{picture}
%\hfill\caption{\label{omega_hartree} The $\omega$-meson contribution
%to the nucleon self energy.}\hfill
%\end{figure}

With the four-Fermi interaction (\ref{4fermii}), there are two distinct
velocity-dependent terms in the quasiparticle interaction, namely the
spatial part of the current-current interaction and the exchange (or Fock)
term of the one-pion-exchange. In the nonrelativistic approximation, their
contributions to $\tilde{F_1}$ are ($\tilde{F_1} = \tilde{F_1^\omega} +
\tilde{F_1^\pi}$) 
\be
\label{fomega}
\tilde{F_1^\omega}&=&\frac{m_N}{m_L^\star} F_1^\omega=
-C_\omega^2\frac{2p_F^3}{\pi^2 m_N^\sigma },\\
\tilde{F_1^\pi}&=& -3\frac{m_N}{p_F}\frac{\der}{\der p}\Sigma_\pi (p)|_{p=p_F},
\ee
respectively. 

Using eq.~(\ref{velo}) we find
\be
(\frac{m_L^\star}{m_N})^{-1}
=\frac{m_N}{m_N^\sigma} +\frac{m_N}{p_F}\frac{\der}{\der p}\Sigma_\pi
(p)|_{p=p_F} = 1-\frac 13 \tilde{F_1}\label{LandauM},
\ee
which implies that
\be
\frac{m_N}{m_N^\sigma}=1-\frac 13 \tilde{F}_1^\omega.\label{Phidefined}
\ee
This formula gives a relation between the $\sigma$-nucleon interaction
(eq.~(\ref{sigmashift})) and the $\omega$-nucleon coupling 
(eq.~(\ref{fomega})). The $\omega$-exchange contribution to the Landau
parameter $F_1$ is due to the velocity-dependent part of the
potential, $\sim \vec{p}_1\cdot\vec{p}_2/m_N^2$.
This is an ${\cal O}(p^2)$ term, and consequently suppressed in naive chiral
counting.  Nonetheless it is this chirally non-leading term in the
four-Fermi interaction (\ref{wform}) that appears on the same footing
with the chirally leading terms in the $\omega$ and $\sigma$ tadpole
graphs. This shows that there must be subtlety in the chiral counting in the
presence of a Fermi sea.

The pion contribution to $F_1$ can be evaluated explicitly \cite{br80}
\be
\frac 13
\tilde{F_1^\pi}&=& -\frac{3f_{\pi NN}^2m_N}{8\pi^2p_F}[\frac{m_\pi^2+2p_F^2}
{2p_F^2} \ln\frac{m_\pi^2+4p_F^2}{m_\pi^2}-2]\nonumber\\
&\approx& -0.153.\label{FockM}
\ee
Here $f_{\pi NN}\approx 1$ is the non-relativistic $\pi$N coupling
constant.
The numerical value of $\tilde{F_1^\pi}$ is obtained at nuclear matter density,
where $p_F\approx 2m_\pi$.

One of the important results of this paper is that
eq.~(\ref{Phidefined}) relates the only unknown
parameter $\tilde{F}_1^\omega$ to the universal scaling factor $\Phi$.
Note that in the absence of the one-pion-exchange interaction -- and
in the nonrelativistic approximation --
$m_N^\sigma$ can be identified with the Landau effective mass
$m_L^\star$. In its presence,
however, the two masses are different due to the pionic Fock term.
We propose to identify the scaling nucleon mass defined in
eq.~(\ref{Nmass}) with the Landau effective mass:
\be\label{mrelation}
m_L^\star=m_N^\star.
\ee
We note that the Landau mass is defined at the Fermi surface, while the
scaling mass refers to a nucleon propagating in a ``vacuum" modified
by the nuclear medium. Although  
the two definitions are closely related, their precise
connection is not understood at present. Nevertheless,
eq.~(\ref{mrelation}) is expected to be a good approximation 
(see also section 5.2).

\section{Orbital Gyromagnetic Ratios in Nuclei}
\indent
Given the effective Lagrangian with the BR scaling and its relation
to Landau Fermi liquid theory,
how can one describe nuclear magnetic moments and axial charge transitions?
This is an important question
because these nuclear processes are sensitive to both the
scaling properties and exchange currents. Here we consider the
gyromagnetic ratios $g_l^{(p,n)}$ of the proton and the neutron in heavy
nuclei, deferring the issue of the nuclear axial-charge transitions
\cite{pmr93} to a later publication \cite{FR}.
We start with the Fermi liquid theory result for the gyromagnetic 
ratio.\footnote{This quantity has been extensively analyzed in terms of
standard exchange currents and their relations, via vector-current
Ward identities, to nuclear forces \cite{riska}.}

\subsection{Migdal's formula}
\indent
The response to a slowly-varying electromagnetic field of an odd
nucleon with momentum $\vec{p}$ added to a closed Fermi sea
can, in Landau theory, be represented by the current \cite{migdal,BWBS}
\be
\vec{J}= \frac{\vec{p}}{m_N}\left(\frac{1+\tau_3}{2} +\frac 16
\frac{F_1^\prime -F_1}{1+F_1/3} \tau_3\right) \label{qpcurrent}
\ee
where $m_N$ is the nucleon mass in medium-free space.
The long-wavelength limit of the current is not unique. 
The physically relevant one
corresponds to the limit $q \rightarrow 0, \omega \rightarrow 0$ with
$q/\omega \rightarrow 0$, where $(\omega,q)$ is the four-momentum transfer.
The current (\ref{qpcurrent}) defines the gyromagnetic ratio
\be
g_l=\frac{1+\tau_3}{2} +\delta g_l
\ee
where
\be
\delta g_l=\frac 16 \frac{F_1^\prime -F_1}{1+F_1/3}\tau_3
=\frac 16 (\tilde{F}_1^\prime-\tilde{F}_1)\tau_3.\label{deltalandau}
\ee

%\begin{figure}[tbh]
%\setlength{\unitlength}{1mm}
%\begin{picture}(150,45)
%\put(75,3){\epsfig{file=current.eps,
%height=35mm,angle=-90}}
%\end{picture}
%\hfill\caption{\label{sp_current} The single particle current.}\hfill
%\end{figure}

\subsection{Chiral Lagrangian results}
\indent
In this section we compute the gyromagnetic ratio using the chiral 
Lagrangian and demonstrate that Migdal's result (\ref{deltalandau}) is
reproduced. The derivation will be made in terms of Feynman diagrams.
The single-particle current $\vec{J}_1=\vec{p}/m_N^\sigma$ is
given by a diagram 
with the external nucleon lines dressed by the scalar and vector fields.
Note that it is the universally scaled mass $m_N^\sigma$ that enters,
not the Landau mass. This leads to a gyromagnetic ratio
\be
(g_l)_{sp}=\frac{m_N}{m_N^\sigma}\frac{1+\tau_3}{2}.\label{SP}
\ee
At first glance this result seems to imply the enhancement of the
single quasiparticle gyromagnetic ratio by the factor $1/\Phi$ 
(for $\Phi <1$) over the free space value. However this interpretation,
often made in the literature, is not correct. We have to take into account
the corrections carefully.

The first correction to (\ref{SP})
is the contribution from short-ranged high-energy
isoscalar vibrations corresponding to an $\omega$ meson.
This contribution has been computed by several authors \cite{matsui,suzuki}.
In the nonrelativistic approximation one finds
\be
g_l^\omega=-\frac 16 C_\omega^2\frac{2p_F^3}{\pi^2}
\frac{1}{m_N^\sigma}=\frac 16 \tilde{F}_1^\omega.\label{deltaomega}
\ee
Now using (\ref{Phidefined}), we obtain the second principal result of
this paper,
\be
g_l^\omega=\frac 16 \tilde{F}_1^\omega= \frac 12
(1-\Phi (\rho)^{-1}).\label{f1phi}
\ee
The corresponding contribution with a $\rho$ exchange in the graph
yields an isovector term
\be
g_l^\rho=-\frac 16 C_\rho^2 \frac{2p_F^3}{\pi^2}\frac{1}{m_N^\sigma}
\tau_3=\frac 16 (\tilde{F}_1^\rho)^\prime \tau_3 \label{deltarho}
\ee
where the constant $C_\rho$ is the coupling strength of the four-Fermi
interaction
\be
\delta \L=-\frac{C_\rho^2}{2} (\bar{N}\gamma_\mu \tau^a N
\bar{N}\gamma^\mu \tau^a N).
\ee
In analogy with the isoscalar channel, we may consider this as arising
when the $\rho$ is integrated out from the Lagrangian, and consequently
identify
\be
C_\rho^2=g_\rho^2/m_\rho^2.
\ee
Again in medium, $m_\rho$ should be replaced by $m_\rho^\star$.
The results (\ref{deltaomega}) and (\ref{deltarho})
can be interpreted in the language of chiral perturbation theory
as arising from four-Fermi interaction counterterms in
the presence of electromagnetic
field, with the counter terms saturated by the $\omega$ and $\rho$ mesons
respectively (see eq.~(92) of \cite{pmr95}).

%\begin{figure}[tbh]
%\setlength{\unitlength}{1mm}
%\begin{picture}(150,60)
%\put(55,5){\epsfig{file=pol_current.eps,
%height=50mm,angle=-90}}
%\end{picture}
%\caption{\label{pol_current} The polarization contribution to the
%current. The backward-going lines correspond to negative energy states
%(anti-nucleons) while the lines without arrows represent nucleon
%states that are blocked by the filled Fermi sea.}
%\end{figure}

The next correction is the pionic exchange current (known as Miyazawa term) 
which yields \cite{br80}
\be
g_l^\pi = \frac 16 ((\tilde{F}_1^\pi)^\prime -\tilde{F}_1^\pi)\tau_3
=-\frac 29 \tilde{F}_1^\pi \tau_3,\label{glpi}
\ee
where the last equality follows from $(\tilde{F}_1^\pi)^\prime = -(1/3)
\tilde{F}_1^\pi$.
Thus, the sum of all contributions is
\be
g_l&=& \frac{m_N}{m_N^\sigma}\frac{1+\tau_3}{2} +\frac 16
(\tilde{F}_1^\omega +(\tilde{F}_1^\rho)^\prime \tau_3)
+ \frac 16 ((\tilde{F}_1^\pi)^\prime -\tilde{F}_1^\pi)\tau_3\nonumber\\
&=&
\frac{1+\tau_3}{2} + \frac 16 (\tilde{F}_1^\prime -\tilde{F}_1)\tau_3
\label{pred}
\ee
where eq.~(\ref{Phidefined}) was used with
\be
\tilde{F}_1 &=& \tilde{F}_1^\omega+\tilde{F}_1^\pi,\label{ftilde}\\
\tilde{F}_1^\prime &=& (\tilde{F}_1^\pi)^\prime +(\tilde{F}_1^\rho)^\prime.
\label{fptilde}
\ee
Thus, when the corrections are suitably calculated, we do
recover the familiar single-particle gyromagnetic ratio $(1+\tau_3)/2$
and reproduce the Fermi-liquid theory result
for $\delta g_l$ (\ref{deltalandau})
\be
\delta g_l =\frac 16 (\tilde{F}_1^\prime-\tilde{F}_1)\tau_3
\label{deltachiral}
\ee
with $\tilde{F}$ and $\tilde{F}^\prime$ in the theory given entirely by
(\ref{ftilde}) and (\ref{fptilde}), respectively.
Equation (\ref{pred}) shows that the isoscalar gyromagnetic ratio is
not renormalized by the medium 
(other than binding effect implicit in the matrix
elements) while the isovector one is. 
{\it It should be emphasized that contrary to naive expectations,
BR scaling is not in conflict with the observed nuclear
magnetic moments.} We will show below that the theory agrees
quantitatively with experimental data.

%\begin{figure}[tbh]
%\setlength{\unitlength}{1mm}
%\begin{picture}(150,50)
%\put(55,5){\epsfig{file=exchangecurr.eps,
%height=40mm,angle=-90}}
%\end{picture}
%\hfill\caption{\label{exchcurr} The pion-exchange current contribution
%to the nucleon current in matter.}\hfill
%\end{figure}
%
%
%
%
\section{Comparison with Experiments}
\subsection{Information from QCD sum rules}
\indent
It is possible to extract the scaling factor $\Phi (\rho)$ from
QCD sum rules -- as well as from an in-medium Gell-Mann-Oakes-Renner
relation \cite{br95} -- and compare with our theory.
In particular, the key information is available from
the calculations of the masses of the $\rho$ meson \cite{sumrules,Jin1}
and the nucleon \cite{CFG,Jin2}
in medium. In their recent
work, Jin and collaborators find (for $\rho=\rho_0$) \cite{Jin1,Jin2}
\be
\frac{m_\rho^\star}{m_\rho}&=& 0.78\pm 0.08,\\
\frac{m_N^\star}{m_N}&=& 0.67\pm 0.05.\label{Jin}
\ee
We identify the $\rho$-meson scaling with the universal
scaling factor,
\be
\Phi (\rho_0)=0.78.\label{sigmaM}
\ee
This is remarkably close to the result that follows from the
GMOR relation in medium \cite{brPR95,CFG2}
\be
\Phi^2 (\rho_0) \approx
\frac{{m_\pi^\star}^2}{m_\pi^2} (1-\frac{\Sigma_{\pi N}\,\rho_0}
{f_\pi^2 m_\pi^2}+\cdots)\approx 0.6,\label{GMOR}
\ee
where the pion-nucleon sigma term $\Sigma_{\pi N}\approx 45$ MeV is used.
In fact, in previous papers by Brown and Rho, the scaling factor
$\Phi$ was inferred from the in-medium GMOR relation.
\subsection{Prediction by chiral Lagrangian}
\indent
Our theory has only one quantity that is not fixed by the theory, namely
the scaling factor $\Phi (\rho)$ ($\tilde{F}_1^\pi$ is of course
fixed for any density by the chiral Lagrangian.).
Since this is given by QCD sum rules
for $\rho=\rho_0$, we use this information to make quantitative prediction.
\subsubsection{Effective nucleon mass}
\indent
The first quantity is the Landau effective mass of the nucleon (\ref{LandauM}),
\be
\frac{m_N^\star}{m_N}&=&\Phi\left(1 + \frac 13 F^\pi_1\right) \nonumber\\
&=& \left(\Phi^{-1}-\frac 13 \tilde{F}_1^\pi\right)^{-1}\nonumber\\
&=& (1/0.78 +0.153)^{-1}=0.69(7)\label{L}
\ee
where we used (\ref{FockM}) and (\ref{sigmaM}).
The agreement with the QCD sum-rule result (\ref{Jin}) is both surprising and
intriguing since as mentioned above, the Landau mass is ``measured" at
the Fermi momentum $p=p_f$ while
the QCD sum-rule mass is defined in the rest frame, so the direct connection
remains to be established.

\subsubsection{Effective axial-vector coupling constant}
\indent
The next quantity of interest is the axial-vector coupling constant
in medium, $g_A^\star$, which can be obtained from the Landau mass
(\ref{LandauM}) and the chiral mass (\ref{Nmass}) as
\be
\frac{g_A^\star}{g_A}=\left(1+\frac 13 F^\pi_1\right)^2=\left(1-\frac 13
\Phi \tilde{F}_1^\pi\right)^{-2},
\ee
which at $\rho=\rho_0$ gives
\be
g_A^\star=1.0(0).
\ee
This agrees well with the observations in heavy nuclei \cite{gAdata}.
Again this is an intriguing result. While it is not understood how this
relation is related to the old one in terms of the Landau-Migdal parameter
$g_0^\prime$ in $NN\leftrightarrow N\Delta$ channel \cite{delta},
it is clearly a short-distance effect in the ``pionic channel"
involving the factor $\Phi$. This supports the argument \cite{pmr93}
that the renormalization of the axial-vector coupling constant in
medium cannot be described in low-order chiral perturbation theory.
\footnote{The $g_A^\star$ calculated here is for a quasiparticle
sitting on top of the Fermi sea and is presumably a fixed-point
quantity as one scales down in the sense of renormalization group flow.
As such, it should be applicable within a configuration
space restricted to near the Fermi surface. I think this is the reason why 
$g_A^\star=1$ was required in the $0\hbar\omega$ Monte Carlo shell-model
calculation of Langanke et al \cite{gAdata}. The consequence of this result
is that if one were to calculate core-polarization contributions
involving multiparticle-multihole configurations mediated by tensor forces,
one should obtain only a minor correction. As Gerry Brown has been arguing for
some time, this can happen because of the suppression of tensor
forces in the presence of BR scaling. Note also that the effective
$g_A^\star$ obtained here has nothing to do with the so-called 
``missing Gamow-Teller strength" often discussed in the literature.} 

\subsubsection{Orbital gyromagnetic ratio}
\indent
Finally, the correction to the single-particle gyromagnetic
ratio can be rewritten as
\be
\delta g_l=\frac 49\left[\Phi^{-1} -1 -\frac 12 \tilde{F}_1^\pi\right]\tau_3
\ee
where we  have used (\ref{glpi}) and the assumption that the nonet relation 
$C_\rho^2=C_\omega^2/9$ holds. The nonet assumption would be justified
if the constants $C_\omega$ and $C_\rho$ were saturated by the $\omega$
and $\rho$ mesons, respectively.
At $\rho=\rho_0$, we find
\be
\delta g_l=0.22(7)\tau_3.\label{deltapred}
\ee
This is in agreement with the result \cite{exp} for protons extracted
from the
dipole sum rule in $^{209}$Bi using the Fujita-Hirata relation \cite{FH}:
\be
\delta g_l^{proton}=\kappa/2 = 0.23\pm 0.03.
\ee
Here $\kappa$ is the enhancement factor in the giant dipole sum rule.
Given that this is extracted from the sum rule in the
giant dipole resonance region, this is a bulk property, so our theory
is directly relevant.

Direct comparison with magnetic moment measurements is difficult
since BR scaling is expected to quench the tensor force which is crucial
for the calculation of contributions from high-excitation states
needed to extract the $\delta g_l$. Calculations
with this effect taken into account are not available at present.
Modulo this caveat, our prediction (\ref{deltapred}) compares well
with Yamazaki's analysis \cite{yamazaki} of magnetic moments
in the $^{208}$Pb region
\be
\delta g_l^{proton} &\approx& 0.33,\nonumber\\
\delta g_l^{neutron}&\approx& -0.22
\ee
and also with the result of Arima et al. \cite{arima,yamazaki}
\be
\delta g_l\approx 0.25\tau_3.
\ee

\subsection*{Acknowledgments}

I am grateful for discussions with Gerry Brown, Begt Friman and
Chaejun Song.
I would also like to thank the organizers of the KOSEF-JSPS Winter School
on ``Recent Developments in Particle and Nuclear Physics" (Seoul, Korea)
for the invitation to lecture at this school.

\end{narrowtext}
\end{document}